\documentclass[preprint,authoryear,12pt]{elsarticle}
\usepackage{graphicx}
\usepackage{a4wide}
\usepackage{epstopdf}
\usepackage{booktabs}
\usepackage{natbib} 
\usepackage{multirow}
\usepackage{amssymb}
 \usepackage{amsthm}
 \usepackage{amsmath}

\journal{Energy Economics}

\begin{document}

\begin{frontmatter}

\title{Commodity futures and market efficiency}

\author[utia,ies]{Ladislav Kristoufek} \ead{kristouf@utia.cas.cz}
\author[utia,ies]{Miloslav Vosvrda} \ead{vosvrda@utia.cas.cz}

\address[utia]{Institute of Information Theory and Automation, Academy of Sciences of the Czech Republic, Pod Vodarenskou Vezi 4, 182 08, Prague, Czech Republic, EU} 
\address[ies]{Institute of Economic Studies, Faculty of Social Sciences, Charles University in Prague, Opletalova 26, 110 00, Prague, Czech Republic, EU}

\begin{abstract}
We analyze the market efficiency of 25 commodity futures across various groups -- metals, energies, softs, grains and other agricultural commodities. To do so, we utilize recently proposed Efficiency Index to find that the most efficient of all the analyzed commodities is heating oil, closely followed by WTI crude oil, cotton, wheat and coffee. On the other end of the ranking, we detect live cattle and feeder cattle. The efficiency is also found to be characteristic for specific groups of commodities -- energy commodities being the most efficient and the other agricultural commodities (formed mainly of livestock) the least efficient groups. We also discuss contributions of the long-term memory, fractal dimension and approximate entropy to the total inefficiency. Last but not least, we come across the nonstandard relationship between the fractal dimension and Hurst exponent. For the analyzed dataset, the relationship between these two is positive meaning that local persistence (trending) is connected to global anti-persistence. We attribute this to specifics of commodity futures which might be predictable in a short term and locally but in a long term, they return to their fundamental price. 

\end{abstract}


\end{frontmatter}

\newpage

\section{Introduction}

Efficient markets hypothesis (EMH) has been a cornerstone of financial economics for decades and it has been brought to the centre by the influential paper of \cite{Fama1970} summarizing empirical findings following the idea of the efficient markets hypothesis by \cite{Fama1965} and \cite{Samuelson1965}. Even though the actual definitions differ, the former study builds on a random walk definition and the latter one on a martingale definition, the qualitative consequences are the same -- the efficiency of a market originates in impossibility of systematic beating of the market, usually in a form of above-average risk-adjusted returns. \cite{Fama1991} later separated the efficiency hypothesis into three forms -- weak, medium and strong -- which differ by different information sets taken into consideration and all are based on inclusion of the information sets in market prices. The weak-form EMH says that all past price movements (and associated statistics) are already reflected in the market prices. Prediction of market movements based on historical time series (technical analysis) is thus not possible for this form. The medium-form EMH states that all publicly available information are already contained in the prices, the strong-form EMH adds all (even privately) available information. The medium-form thus discards fundamental analysis as well and the strong-form eliminates even insiders from making profit. Evidently, a weaker form of EMH is always a subset of a stronger form. Even though EMH has been repeatedly disparaged both empirically and theoretically \citep{Cont2001,Malkiel2003}, and even more so after an outbreak of the Global Financial Crisis in 2007/2008, its validity remains an open issue, yet still it persists in standard textbooks of financial economics \citep{Elton2003}.

Comparison of efficiency across various assets has been discussed in different studies. In a series of papers, \cite{DiMatteo2003,DiMatteo2005} and \cite{DiMatteo2007} study long-term memory and multi-scaling of a wide portfolio of stock indices, foreign exchange rates, Treasury rates and Eurodollar interbank interest rates using various estimators of long-term memory. They show that stock indices of more developed countries are also more efficient yet showing a weak signs of anti-persistence (properties of long-term memory are described in detail in the Methodology section), finding no deviations from EMH for all analyzed maturities of Eurodollar and Treasury rates. For US dollar exchange rates, the authors find diverse results with no evident pattern connecting exchange rate efficiency level with geographical or geopolitical properties. In another series of papers, \cite{Cajueiro2004b,Cajueiro2004,Cajueiro2004a,Cajueiro2005} compare stock market indices from different continents finding that the US and Japanese markets are the most efficient ones whereas the Asian and Latin American ones are detected as the least efficient ones. \cite{Lim2007} studies non-linear dependencies, their evolution in time and connection to market efficiency for a set of stock markets. The author finds the US market to be the most efficient one followed by Korea, Taiwan and Japan. On the other end of the ranking, Malaysia, Chile and Argentina are placed. \cite{Zunino2010} utilize the complexity-entropy causality plane to rank stock market indices to show that the emergent markets are less efficient than the developed ones as one would expect. The difference is attributed to a lower entropy and a higher complexity of the emergent markets. \cite{Kristoufek2013} introduce the Efficiency Index and come up with a ranking of stock market indices finding that the most efficient markets are located in Western Europe, USA and Japan whereas the least efficient markets are situated in Latin America and Asia.

However, up to our best knowledge, proper attention has not been given to a comparison of the efficiency of commodity markets. In this paper, we analyze futures markets for a wide range of commodities -- energy, metals and various agricultural commodities -- and compare their efficiency using the Efficiency Index proposed by \cite{Kristoufek2013}. The paper is structured as follows. Section 2 covers literature dealing with the efficiency of commodities. Section 3 describes the methodology in detail. Section 4 describes the analyzed dataset and brings the results. Section 5 concludes. We show that efficiency is related to a type of commodity (energy commodities being the most efficient ones and other agricultural commodities being the least efficient ones). In addition, we find a nonstandard relationship between the local and global properties of the series as most of the series show local persistence while in global, they are mean-reverting. The series thus follow quite strong local trends but in a long term, they return to their fundamental value.

\section{Literature review} 

Testing the market efficiency in commodities markets has a long history. \cite{Roll1972} examines the commodity price index and argues that the market is inefficient due to significant serial correlations of its returns. \cite{Danthine1977} disputes such claim and shows that the violation of the standard martingale condition does not imply inefficiency in the commodity spot markets with support of risk aversion and no arbitrage opportunities. \cite{Gjolberg1985} analyzes oil spot prices at the Rotterdam market, rejects the efficiency hypothesis and constructs a profitable trading rule for daily, weekly and monthly price changes. \cite{Panas1991} studies the Rotterdam oil market as well together with the Italian market and based on leptokurtic monthly price changes, he rejects the markets' efficiency. \cite{Herbert1996} examine the US spot (cash) and futures markets for natural gas and find these to be inefficient. They argue that such inefficiency is caused by the specific structure of the US gas markets. 

More recently, \cite{Tabak2007} analyze the efficiency of Brent and WTI crude oil using the rescaled range analysis and show that the markets are becoming more efficient in time. \cite{Alvarez-Ramirez2008} study the auto-correlation structure of the crude oil process using the detrended fluctuation analysis. They show that in long-term, the market is efficient but in short-term, the auto-correlation structure leads to rejection of the efficiency. \cite{Alvarez-Ramirez2010} further inspect the crude oil markets using lagged detrended fluctuation analysis and argue that multi-scaling and deviations from the random walk behavior cause the spot prices to be inefficient. The research on evolution of efficiency in time is further extended by \cite{Wang2010} where the authors study short-, medium- and long-term efficiency for various scales of the detrended fluctuation analysis approach. They show that the WTI crude oil becomes more efficient in time for all three analyzed scales. Using also the detrended fluctuation analysis, \cite{Wang2011} argue that WTI crude oil spot and futures are not efficient for short time scales below one month. Crude oil markets (Brent and WTI) are also analyzed by \cite{Charles2009} who use the variance ratio tests to show that the Brent market is weak-form efficient but the WTI market is not while providing some discussion about effects of deregulation on the markets.

\cite{Lee2009} study four energy commodities -- coal, oil, gas and electricity -- using panel data stationarity tests to uncover that none of the studied markets is efficient in the strict stationarity sense. \cite{Lean2010} study WTI crude oil spot and futures prices using mean-variance and stochastic dominance approaches finding no arbitrage opportunities between spot and futures prices while the findings are robust for various sub-periods and critical events. \cite{Narayan2010} study long-term relationship between spot and futures prices of gold and oil. They find that investors use the gold market to hedge against inflation and for our purposes also more importantly that crude oil market predicts the gold market and vice versa implying inefficiency.

\cite{Wang2010a} study high-frequency futures data of crude oil, heating oil, gasoline and natural gas using various nonlinear models. For heating oil and natural gas, the authors find market inefficiencies which are profound mainly during the bull market conditions. \cite{Gebre-Mariam2011} focuses on the US natural gas market (spot and futures) finding no arbitrage opportunities for daily prices but in general, the author claims that the markets can be seen as efficient only for contracts with approximately a month to maturity. \cite{Martina2011} utilize entropy approaches to WTI crude oil spot prices and find various cycles in its prices. Entropy is also applied by \cite{Ortiz-Cruz2012} who again study daily WTI prices finding the market to be efficient with two episodes of inefficiency connected to the early 1990s and late 2000s US recessions. The authors stress that deregulation of the market has helped improving its efficiency. 

\cite{Zunino2011} apply information theory methods (specifically the permutation entropy and permutation statistical complexity) to the commodity markets allowing them for efficiency ranking finding silver, copper and cotton to be the most efficient commodities. \cite{Wang2011} study the gold market using the multifractal detrended fluctuation analysis to show that the market becomes more efficient in time especially after 2001. \cite{Kim2011} use the random matrix theory and network analysis to show that stock and commodity markets are well decoupled except for oil and gold showing signs of inefficiency. \cite{Kim2011a} then focus on the Korean agricultural market using the detrended fluctuation analysis finding anti-correlated series with strong volatility clustering aiming at inefficiency. 

Out of these selected papers, it is evident that analysis of efficiency of commodity markets is fruitful with many approaches to the topic. However, the studies usually focus on a single (or a pair) of efficiency measures to test whether the specific markets are or are not efficient. Moreover, the analysis is usually strongly focused on a single commodity or a small group of commodities. Here, we contribute to the literature by applying various efficiency measures on a wide portfolio of commodities ranging from energy and agricultural (with several subgroups) commodities to metals. Moreover, we utilize the efficiency measure introduced by \cite{Kristoufek2013} to rank the commodities according to their efficiency.

\section{Methodology}

Efficient market can be defined in several ways. The main distinction roots back to 1965 when \cite{Fama1965} and \cite{Samuelson1965} used different definitions -- a random walk and a martingale, respectively. We stick to the martingale definition efficiency because it is less restrictive. Based on this definition, we assume that the returns of a financial asset are serially uncorrelated and with finite variance for the efficient market situation. Such a simple definition allows to use various measures of market efficiency, which are described in this section. Eventually, we refer to the Efficient Index which takes these statistics into consideration and it helps to rank different assets according to their efficiency while using various dynamic properties of the time series under study.

\subsection{Long-term memory}

Long-term memory (long-range dependence) series are characteristic with values in (even distant, in theory infinitely distant) past influencing the present and future values. These processes are standardly described with the long-term memory parameter $H$ (Hurst exponent) which ranges between $0 \le H <1$ for stationary invertible processes. The midpoint, $H=0.5$, holds for uncorrelated (or in general short-term memory) processes, i.e. processes of the efficient market. For $H>0.5$, the processes are positively correlated with long-term memory and are usually referred to as persistent. These processes systematically follow local trends while still remaining stationary. For $H<0.5$, we have long-term memory processes with negative correlations -- anti-persistent processes. Such processes switch the direction more often than a random process does.

More formally, the long-term memory processes are defined in both time and frequency domains. In the time domain, it is connected to a power-law decaying auto-correlation function. For the auto-correlation function $\rho(k)$ with time lag $k$, the decay is characterized as $\rho(k) \propto k^{2H-2}$ for $k\rightarrow +\infty$. In the frequency domain, the spectrum $f(\lambda)$ with frequency $\lambda$ of the long-range dependent process diverges at the origin so that $f(\lambda)\propto \lambda^{1-2H}$ for $\lambda \rightarrow 0+$. These definitions further lead to non-summable auto-correlations and diverging covariance of partial sums of the process for the persistent series. These properties are used in various estimators of parameter $H$. For comparison of both time and frequency domain estimators, see \cite{Beran1994,Taqqu1995,Taqqu1996,Robinson1995a,Geweke1983,DiMatteo2003,DiMatteo2007,Barunik2010,Teverovsky1999}. Out of theses estimators, we opt for the local Whittle and GPH estimators which are suitable for short time series with a possible weak short-term memory \citep{Taqqu1995,Taqqu1996}, which can easily bias the time domain estimators \citep{Teverovsky1999,Kristoufek2012}. Moreover, these estimators have well-defined asymptotic properties -- they are consistent and asymptotically normal estimators.

\subsubsection*{Local Whittle estimator}
The local Whittle estimator \citep{Robinson1995a} is a semi-parametric maximum likelihood estimator utilizing a likelihood function of \cite{Kunsch1987} and focusing only on a part of the spectrum $f(\lambda)$ near the origin. The full parametric specification is thus not needed and one does not need to assume any specific underlying long-term memory model but only a model with divergent at origin spectrum. This way, the estimator does not take into consideration high frequencies and it is in turn resistant to the short-term memory bias. As an estimator of the spectrum of series $\{x_t\}$, we use the periodogram defined as $I(\lambda_j)=\frac{1}{T}\sum_{t=1}^{T}{\exp(-2\pi i t\lambda_j)x_t}$ with $j=1,2,\ldots,m$ where $m\le T/2$ and $\lambda_j=2\pi j/T$. The local Whittle estimator is defined as
\begin{equation}
\label{eq:LWX}
\widehat{H}=\arg \min_{0\le H <1} R(H),
\end{equation} 
where 
\begin{equation}
\label{eq:LWX_R}
R(H)=\log\left(\frac{1}{m}\sum_{j=1}^m{\lambda_j^{2H-1}I(\lambda_j)}\right)-\frac{2H-1}{m}\sum_{j=1}^m{\log \lambda_j}.
\end{equation}
The local Whittle estimator is consistent and asymptotically normal, specifically 
\begin{equation}
\sqrt{m}(\widehat{H}-H^0) \rightarrow_d N(0,1/4).
\end{equation}

\subsubsection*{GPH estimator}

Contrarily to the local Whittle estimator, the GPH estimator, named after the authors of \cite{Geweke1983}, is based on a full functional specification of the underlying process as the fractional Gaussian noise implying a specific spectral form:
\begin{equation}
\log f(\lambda) \propto -(H-0.5)\log (4\sin^2(\lambda/2))
\end{equation}
Again, the spectrum is estimated using the periodogram and the Hurst exponent is estimated using the ordinary least squares on
\begin{equation}
\label{GPH}
\log I(\lambda_j) \propto -(H-0.5)\log (4\sin^2(\lambda_j/2)).
\end{equation}
The GPH estimator is consistent and asymptotically normal \citep{Beran1994}, specifically 
\begin{equation}
\sqrt{T}(\widehat{H}-H^0) \rightarrow_d N(0,\pi^2/6).
\end{equation}

The GPH estimator is thus asymptotically infinitely more efficient than the local Whittle estimator. However, this is true only if the true underlying process is in fact the fractional Gaussian noise. In financial and economic time series, this is frequently not the case as the processes are mostly a combination of short-term (such us autoregressive moving average -- ARMA -- processes of various specifications) and long-term memory (such as the aforementioned fractional Gaussian noise of fractionally integrated ARMA) processes. In this case, the GPH estimator becomes biased. To avoid the bias, the GPH estimator is based only on a part of the spectrum (periodogram) close to the origin as for the local Whittle estimator. The regression in Eq. \ref{GPH} is then applied only for a part of the periodogram based on the same parameter $m$ as for the local Whittle estimator\footnote{In our analysis, we apply $m=T^{0.6}$ \citep{Phillips2004}.}.

\subsection{Fractal dimension}

Contrary to the long-term memory, which can be seen as a characteristic of global dependence and correlation structure, the fractal dimension $D$ can be taken as a measure of local memory of the series as it is a measure of roughness of the series \citep{Kristoufek2013}. As the series can be differently rough or smooth for its specific parts, it can be locally serially correlated even though on a global level, the correlations might vanish and are not necessarily observable or detectable. 

For a univariate series, the fractal dimension ranges between $1<D\le 2$. For self-similar processes, the fractal dimension is tightly connected to the Hurst exponent (long-term memory) of the series so that $D=2-H$. In economic terms, this can be understood as a perfect transmission of a local behavior (fractal dimension) to a global behavior (long-term memory). However, the relation usually does not hold perfectly for the financial series so that both $D$ and $H$ give different insights into the dynamics of the series making it worth studying them separately. 

In general, $D=1.5$ holds for an uncorrelated series with no local trending or no local anti-correlations and thus it is also a value of $D$ for the efficient market. For a low fractal dimension $D<1.5$, the roughness of the series is lower than for an uncorrelated process so that we observe local trending and the series is said to be locally persistent. Reversely, a high fractal dimension $D>1.5$ characterizes a series rougher than the uncorrelated one, which is connected to local anti-persistence, i.e. the series are negatively auto-correlated locally. For purposes of the Efficiency Index introduced later in this section, we utilize Hall-Wood and Genton estimators \citep{Gneiting2004,Gneiting2010}.

\subsubsection*{Hall-Wood estimator}

Hall-Wood estimator \citep{Hall1993} is a box-counting procedure which utilizes scaling of absolute deviations between steps. Formally, we have
\begin{equation}
\widehat{A(l/n)}=\left\lfloor\frac{l}{n}\right\rfloor\sum_{i=1}^{\lfloor n/l \rfloor}{|x_{i \lfloor l/n \rfloor}-x_{(i-1)\lfloor l/n \rfloor}|}
\end{equation}
representing the absolute deviations. Using the definition of the fractal dimension \citep{Gneiting2004,Gneiting2010}, the Hall-Wood estimator is given by
\begin{equation}
\widehat{D_{HW}}=2-\frac{\sum_{l=1}^{L}{(s_l-\bar{s})\log(\widehat{A(l/n)})}}{\sum_{l=1}^{L}{(s_l-\bar{s})^2}}
\end{equation}
where $L \ge 2$, $s_l=\log(l/n)$ and $\bar{s}=\frac{1}{L}\sum_{l=1}^{L}{s_l}$. To minimize potential bias, \cite{Hall1993} propose using $L=2$ so that we obtain the estimate of the fractal dimension $\widehat{D_{HW}}$ as
\begin{equation}
\widehat{D_{HW}}=2-\frac{\log\widehat{A(2/n)}-\log\widehat{A(1/n)}}{\log2}.
\end{equation}

\subsubsection*{Genton estimator}

\cite{Gneiting2004} and \cite{Gneiting2010} propose a method of moments estimator based on the robust variogram of \cite{Genton1998}. The variogram is defined as
\begin{equation}
\widehat{V_2(l/n)}=\frac{1}{2(n-l)}\sum_{i=l}^{n}{(x_{i/n}-x_{(i-l)l/n})^2},
\end{equation}
and the Genton estimator is obtained as
\begin{equation}
\widehat{D_{G}}=2-\frac{\sum_{l=1}^{L}{(s_l-\bar{s})\log(\widehat{V_2(l/n)})}}{2\sum_{l=1}^{L}{(s_l-\bar{s})^2}}
\end{equation}
where again $L \ge 2$, $s_l=\log(l/n)$ and $\bar{s}=\frac{1}{L}\sum_{l=1}^{L}{s_l}$. \cite{Davies1999} and \cite{Zhu2002} again suggest to use $L=2$ to reduce the potential bias so that the estimate $\widehat{D_{G}}$ reads
\begin{equation}
\widehat{D_{G}}=2-\frac{\log\widehat{V_2(2/n)}-\log\widehat{V_2(1/n)}}{2\log2}.
\end{equation}

\subsection{Approximate entropy}

Entropy can be considered as a measure of complexity of the considered system. The systems with high entropy can be characterized by no information flows and are thus random up to uncertainty and reversely, the series with low entropy can be seen as deterministic \citep{Pincus2004}. The efficient market can be then seen as the one with maximum entropy and the lower the entropy, the less efficient the market is. For purposes of the Efficiency Index, we need an entropy measure which is bounded. Therefore, we utilize the approximate entropy introduced by \cite{Pincus1991}.

Let $m$ be a positive integer and let $r$ be a positive real number. For a time series $\left\{ u_{1},u_{2},...,u_{T}\right\}$, with a time series length $T$, let us form a sequence of vectors $\mathbf{X}_{1}, \mathbf{X}_{2},...,\mathbf{X}_{T-m+1}$ in $\mathbf{R}^{m}$ where $\mathbf{X}_{i}=\left( u_{i},u_{i+1},...,u_{i+m-1}\right)$. Using the Takens metrics of distance
\begin{equation}
d\left[ \mathbf{X}_{i},\mathbf{X}_{j}\right] =\underset{k=1,...,m}{\max }\left( \left\vert u\left( i+k-1\right) -u\left(j+k-1\right) \right\vert \right),
\end{equation}
and defining a characteristic function $\chi_{i}^{m}(r)$ as a number of times $d\left[ \mathbf{X}_{i},\mathbf{X}_{j}\right] \leq r/\left( T-m+1\right)$ for each $1\leq i\leq N-m+1$, we define
\begin{gather}
\Phi ^{m}(r) = \frac{1}{T-m+1} \sum_{i=1}^{T-m+1}\log \left[ \chi _{i}^{m}(r) \right]
\end{gather}
which is further used in
\begin{gather}
ER_{m}=\underset{r\rightarrow 0}{\lim }\underset{T\rightarrow \infty }{\lim }\left[ \Phi ^{m}(r) -\Phi ^{m+1}(r) \right].
\end{gather}
The approximate entropy (ApEn) is then defined as
\begin{gather}
ApEn=\underset{m\rightarrow \infty }{\lim }ER_{m}.
\end{gather}
Since $r$ can be seen as an discriminating factor for the distance measured by the Takens metrics and $m$ is the number of elements whose closeness is measured, the approximate entropy measures whether different segments of the series follow similar patterns. For an identically uniformly independently distributed random process, the approximate entropy converges to $-\log \left( r/\sqrt{3}\right) $ for all $m$ \citep{Pincus1991}. For a completely deterministic process, the entropy goes to 0. Therefore, we can rescale the approximate entropy so that $0 \le ApEn \le 1 $, where 0 characterizes a completely deterministic process and 1 a completely uncertain process characteristic for the efficient market. In turn, it can be utilized in the Efficiency Index, definition of which follows.

\subsection{Capital market efficiency measure}

\cite{Kristoufek2013} introduce the Efficiency Index (EI) is defined as

\begin{equation}
EI=\sqrt{{\sum_{i=1}^n{\left(\frac{\widehat{M_i}-M_i^{\ast}}{R_i}\right)^2}}},
\end{equation}
where $M_i$ is the $i$th measure of efficiency, $\widehat{M_i}$ is an estimate of the $i$th measure, $M_i^{\ast}$ is an expected value of the $i$th measure for the efficient market and $R_i$ is a range of the $i$th measure. In words, $EI$ is simply a distance from the efficient market situation. Here, we base the index on three measures of market efficiency -- Hurst exponent $H$ with an expected value of 0.5 for the efficient market ($M_H^{\ast}=0.5$), fractal dimension $D$ with an expected value of 1.5 ($M_D^{\ast}=1.5$) and the approximate entropy with an expected value of 1 ($M_{AE}^{\ast}=1$). Hurst exponent is taken as an average of the GPH and the local Whittle estimates. In the same way, the fractal dimension is set as an average of the Hall-Wood and Genton estimates. For the approximate entropy, we utilize the estimate described in the corresponding section. The approximate entropy need to be rescaled as it ranges between 0 and 1 with the efficient market of $ApEn=1$. We thus have $R_{AE}=2$ and $R_D=R_H=1$.

\section{Data description and results}

We analyze daily front futures prices, i.e. futures with the earliest delivery, of 25 commodities in period between 1.1.2000 and 22.7.2013\footnote{The time series were obtained from http://www.quandl.com server on 23.7.2013.}. The dataset contains 4 energy (Brent crude oil, WTI crude oil, heating oil and natural gas), 5 metals (copper, gold, silver, palladium and platinum), 7 grains (corn, oats, rough rice, soybean meal, soybean oil, soybeans and wheat), 5 softs (cocoa, coffee, cotton, orange juice and sugar) and 4 other agricultural commodities (feeder cattle, lean hogs, live cattle and lumber) futures from Chicago Board of Trade (CBOT), Chicago Mercantile Exchange (CME), IntercontinentalExchange (ICE), New York Mercantile Exchange (NYMEX) and its division Commodity Exchange (COMEX), which are summarized in Tab. \ref{tab1}. We analyze logarithmic prices $S_{i,t}=\log P_{i,t}$, where $P_{i,t}$ is the price of futures $i$ at time $t$, for the fractal dimension and logarithmic returns $r_{i,t}=S_{i,t}-S_{i,t-1}$ for the long-term memory and approximate entropy. The returns of all the analyzed futures are stationary according to ADF \citep{Dickey1979} and KPSS \citep{Kwiatkowski1992} tests (we do not report the $p$-values here).

Estimated Hurst exponents, fractal dimensions and approximate entropies are summarized in Tab. \ref{tab2}. We observe that majority of commodities is characteristic with the fractal dimension below 1.5 which indicates local persistence. These series are thus locally trending. This is most evident for feeder cattle, lean hogs and live cattle, i.e. majorly livestock futures. On the contrary, the energy commodities -- namely both the crude oils and natural gas -- are close to fractal dimension of 1.5 and as such, they do not show any signs of local inefficiencies. For the long-term memory part, most of the futures are below 0.5 indicating anti-persistence which translates into a mean-reversion of prices, something that is not standardly observed for stocks, stock indices and exchange rates which are characteristic by a unit-root behavior. The strongest anti-persistence is seen for cocoa, oats and orange juice. Nonetheless, there is a portion of commodities which show signs of persistence. These are copper, palladium, platinum and sugar. Cotton and natural gas get the closest to the value of the efficient market. For the approximate entropy, several values are close to 1 for the efficient market\footnote{Several values even reach value above 1 due to the finite sample.} -- lumber, sugar and heating oil. The most complex, and thus the least efficient, series include feeder cattle and live cattle.

\begin{figure}[htbp]
\center
\begin{tabular}{c}
\includegraphics[width=6in]{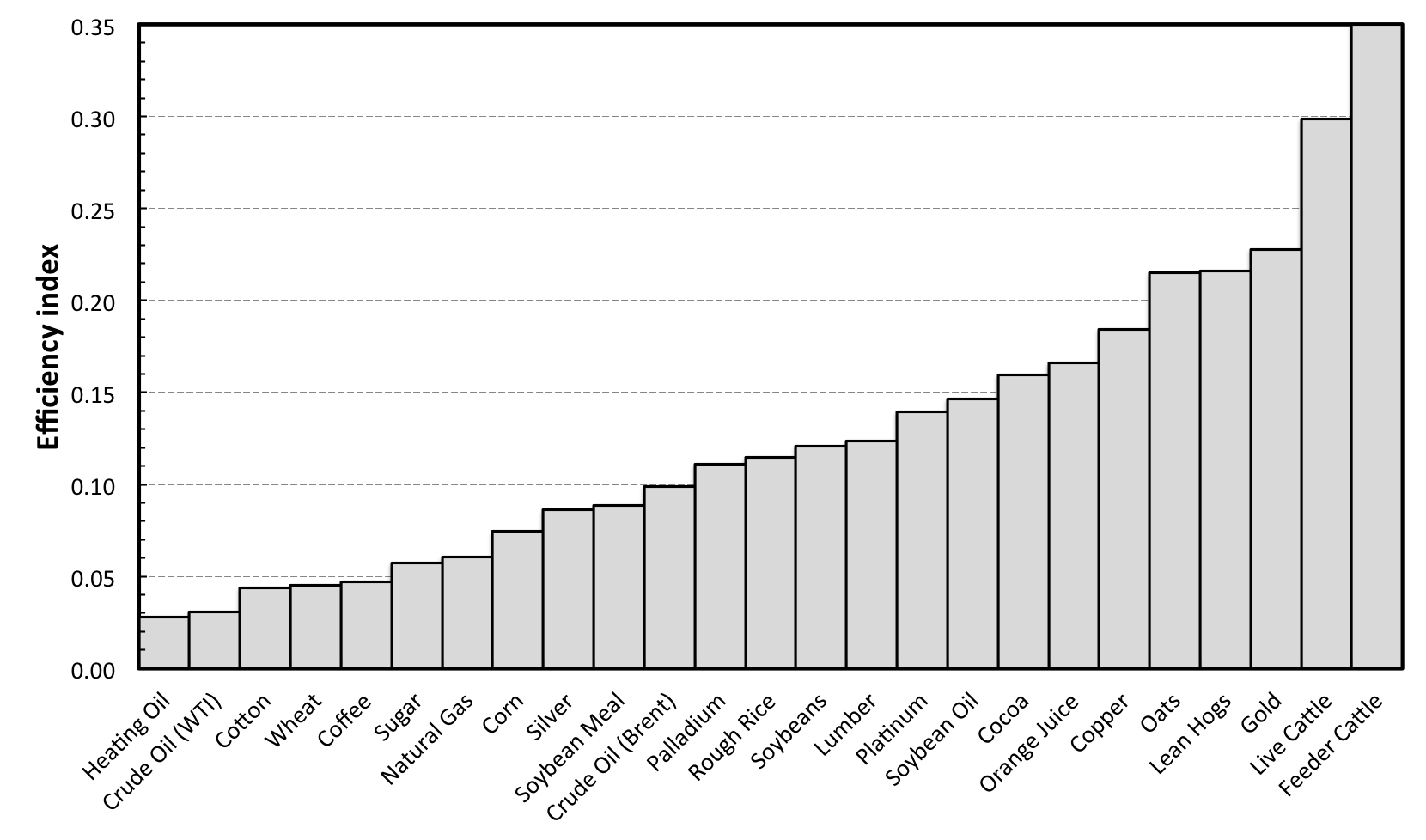}\\
\end{tabular}
\caption{\footnotesize\textbf{Efficiency Index of commodity futures.} Commodities are sorted from the most efficient one (left) to the least efficient one (right).\label{EI}}
\end{figure}

\begin{figure}[htbp]
\center
\begin{tabular}{c}
\includegraphics[width=6in]{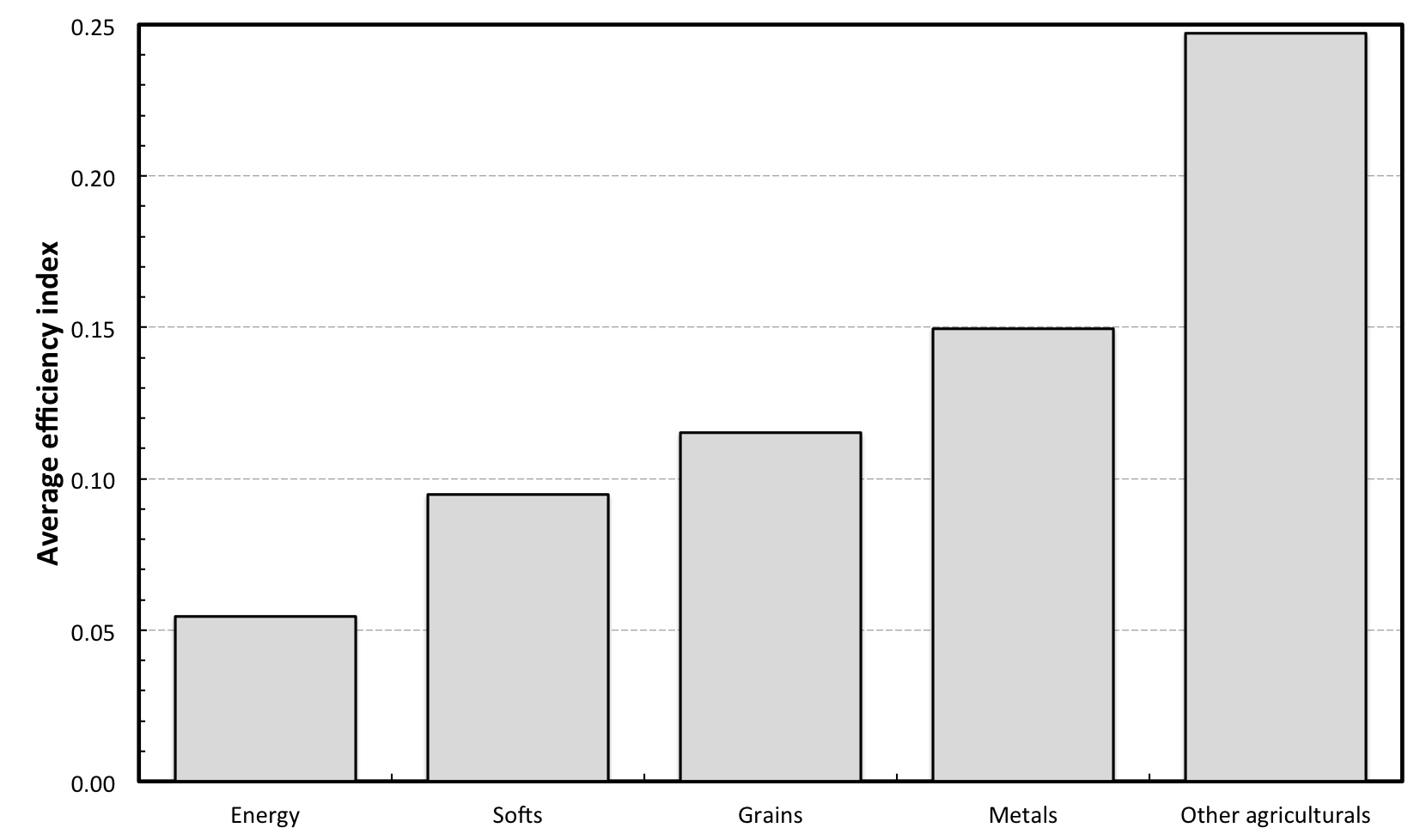}\\
\end{tabular}
\caption{\footnotesize\textbf{Average Efficiency Index for groups of commodities.} Groups are sorted from the most (left) to the least (right) efficient ones.\label{Means}}
\end{figure}

Putting these results together, we arrive at the Efficiency Indices and efficiency ranking which are graphically represented in Fig. \ref{EI}. The most efficient of the commodities turns out to be heating oil closely followed by WTI crude oil. Cotton, wheat and coffee come after these with a similar level of efficiency. The ranking is then supplemented by other commodities, efficiency of which increases quite steadily across the ranking. Feeder cattle is the least efficient commodity in this set quite closely followed by live cattle. The livestock futures thus seem to be rather inefficient compared to the others. Connected to this finding, we also show an average Efficiency Index for commodities according to their type. In Fig. \ref{Means}, we can see that the energy futures are the most efficient ones followed by softs, grains and metals. By far the least efficient group consists of the other agricultural commodities, i.e. feeder cattle, lean hogs, live cattle and lumber. This is well in hand with the observations about very inefficient livestock futures.

\begin{figure}[htbp]
\center
\begin{tabular}{c}
\includegraphics[width=6in]{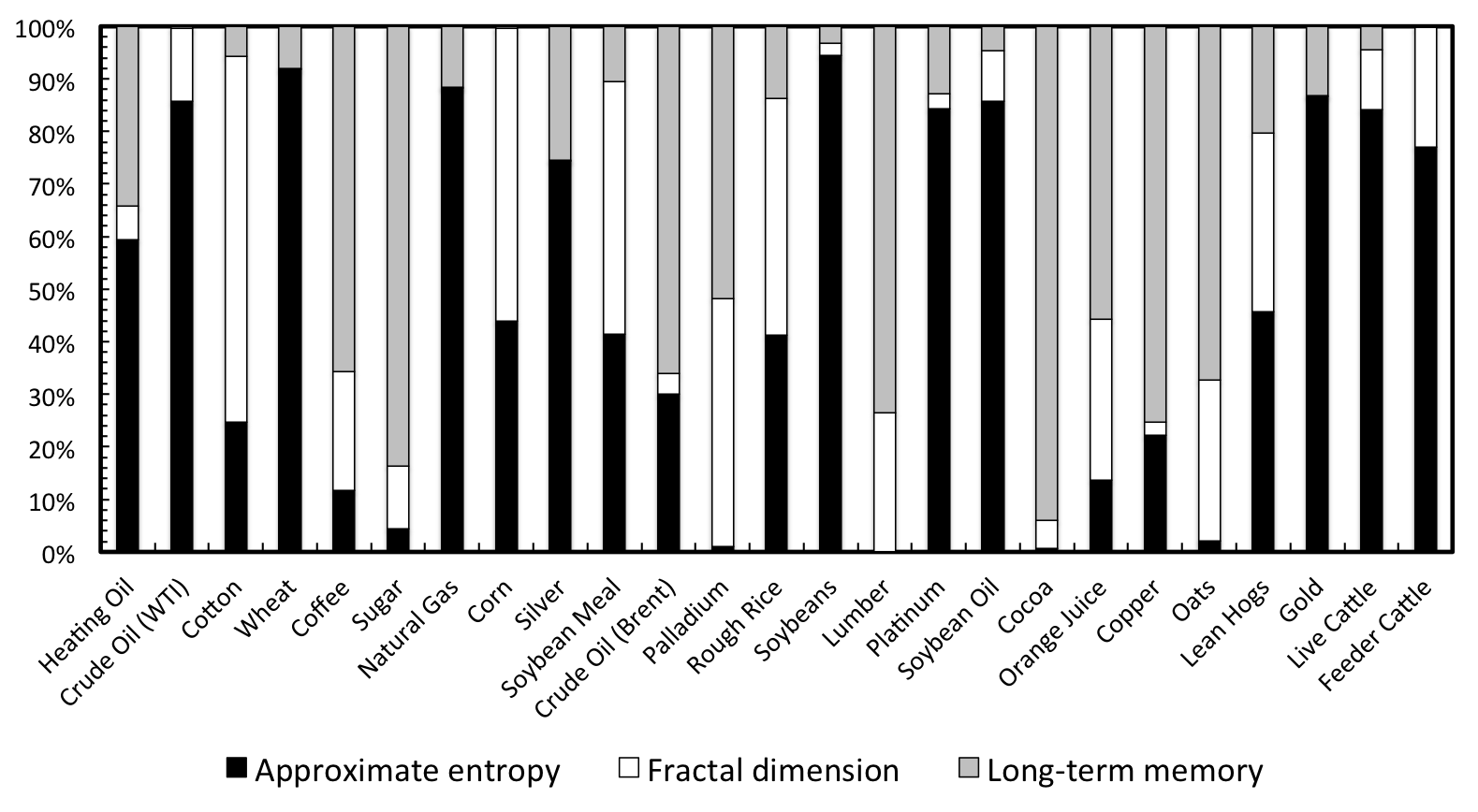}\\
\end{tabular}
\caption{\footnotesize\textbf{Contribution to inefficiency I.} Commodities are sorted according to their efficiency with respect to Fig. \ref{EI}.\label{Parts1}}
\end{figure}

\begin{figure}[htbp]
\center
\begin{tabular}{c}
\includegraphics[width=6in]{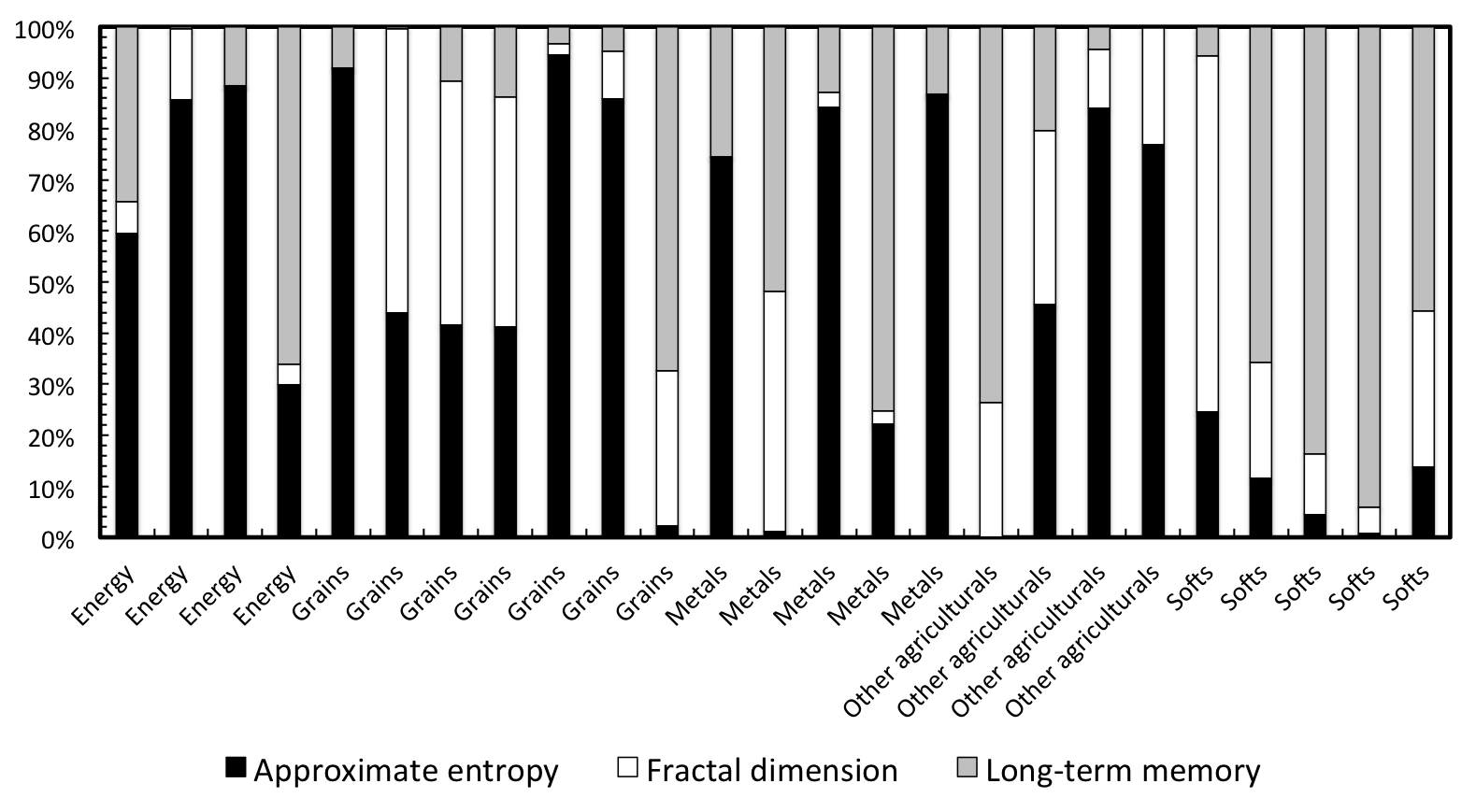}\\
\end{tabular}
\caption{\footnotesize\textbf{Contribution to inefficiency II.} Commodities are sorted according to their group.\label{Parts2}}
\end{figure}

In Figs. \ref{Parts1} and \ref{Parts2}, we decompose the efficiency index into its parts. In Fig. \ref{Parts1}, the actual futures ranked according to the Efficiency Index are represented, and in Fig. \ref{Parts2}, these are sorted according to their type to better see possible patterns and regularities. We observe that for about half of the futures, the approximate entropy is the dominant inefficiency source. Interestingly, it is the most important part for both the most and the least efficient commodities. For the others, the long-term memory part is dominant. Fractal dimension forms usually only a smaller part of the inefficiency and only for wheat, it contributes the most. When we look at the whole groups of commodities, we observe that for energy, grains and other agricultural commodities, the approximate entropy forms an important or even a dominant part for most of them. For grains, fractal dimension creates a significant part for three of the group. For softs, the long-term memory is the most important of the inefficiency contributors. And for metals, the evidence is mixed.

\begin{figure}[htbp]
\center
\begin{tabular}{c}
\includegraphics[width=6in]{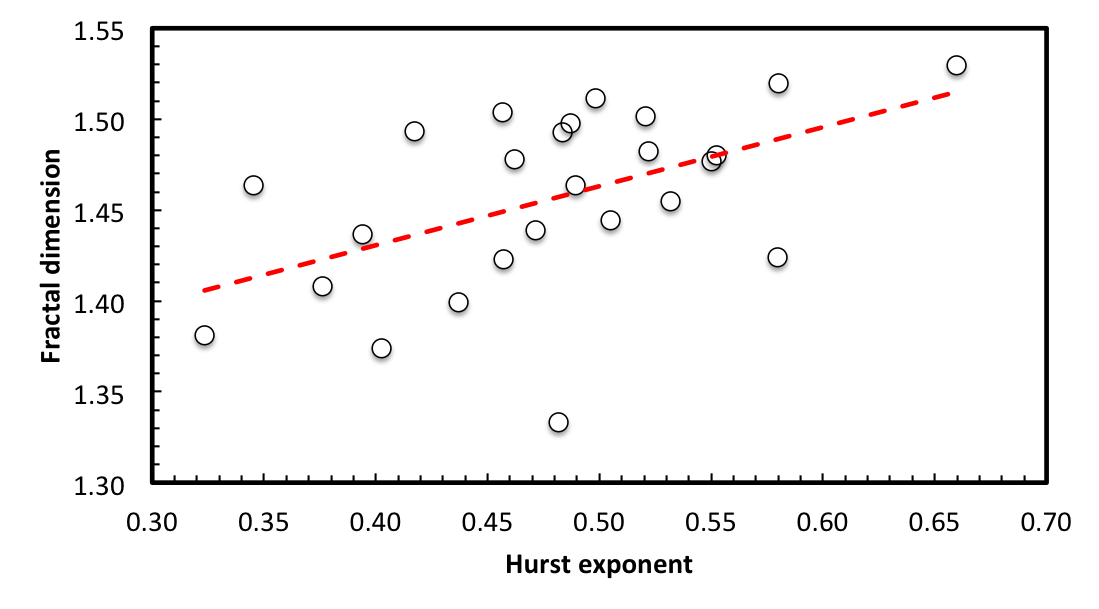}\\
\end{tabular}
\caption{\footnotesize\textbf{Relationship between Hurst exponent and fractal dimension.} For self-similar processes, we expect $D=2-H$, i.e. a negative slope. The red dashed line represents the least squares fit uncovering positive relationship between $D$ and $H$.\label{DH}}
\end{figure}

Fig. \ref{DH} then illustrates a relationship between fractal dimension and Hurst exponent. For self-similar processes, it holds that $D=2-H$. In economic terms, self-similar processes are characteristic by translating the local properties into the global ones. Therefore, for a locally persistent process with $D<1.5$, this translates into the global persistence with $H>0.5$, and vice versa. However, we do not observe such relationship for the analyzed commodities. Actually, the dependence is reversed so that with the increasing Hurst exponent, the fractal dimension increases. This is in contrast with the results for stock indices \citep{Kristoufek2013}. Nonetheless, such result can be well explained by characteristics of commodities futures -- locally (or in the short term), the changes in futures prices are partially predictable, but globally, the prices return to their fundamental values. 

\section{Conclusion}

We have analyzed the market efficiency of 25 commodities futures across various groups -- metals, energies, softs, grains and other agricultural commodities. To do so, we have utilized the recently proposed Efficiency Index to find that the most efficient of all the analyzed commodities is heating oil, closely followed by WTI crude oil, cotton, wheat and coffee. On the other end of the ranking, we have detected live cattle and feeder cattle. The efficiency also seems to be characteristic for specific groups of commodities -- energy commodities have been found the most efficient, followed by softs, grains and metals whereas the other agricultural commodities (formed mainly of livestock) form the least efficient group. Apart from that, we have also discussed the contributions of the long-term memory, fractal dimension and approximate entropy to the total inefficiency. We have uncovered that the contribution is type-dependent as well even though the regularities are not strongly pronounced. Last but not least, we have come across the nonstandard relationship between the fractal dimension and Hurst exponent. For the analyzed dataset, the relationship between these two is positive meaning that local persistence (trending) is connected to global anti-persistence. We attribute this to specifics of commodity futures which might be predictable in a short term and locally but in a long term, they return to their fundamental price, which differs from the results found for stock indices \citep{Kristoufek2013}. 

\section*{Acknowledgements}
 
The support from the Czech Science Foundation under Grants 402/09/0965 and P402/11/0948, and project SVV 267 504 are gratefully acknowledged.

\section*{References}
\bibliography{EI_EE}
\bibliographystyle{chicago}

\newpage

\begin{table}[htbp]
\centering
\caption{Analyzed commodities}
\label{tab1}
\footnotesize
\begin{tabular}{c|c|c}
\toprule \toprule
Full name&Short name&Type\\
\midrule \midrule
CBOT Corn C1&Corn&Grains\\
CBOT Oats O1&Oats&Grains\\
CBOT Rough Rice RR1&Rough Rice&Grains\\
CBOT Soybean Meal SM1&Soybean Meal&Grains\\
CBOT Soybean Oil BO1&Soybean Oil&Grains\\
CBOT Soybeans S1&Soybeans&Grains\\
CBOT Wheat W1&Wheat&Grains\\
CME Feeder Cattle FC1&Feeder Cattle&Other agriculturals\\
CME Lean Hogs LN1&Lean Hogs&Other agriculturals\\
CME Live Cattle LC1&Live Cattle&Other agriculturals\\
CME Lumber LB1&Lumber&Other agriculturals\\
COMEX Copper HG1&Copper&Metals\\
COMEX Gold GC1&Gold&Metals\\
COMEX Silver SI1&Silver&Metals\\
ICE Brent Crude Oil B1&Crude Oil (Brent)&Energy\\
ICE Cocoa CC1&Cocoa&Softs\\
ICE Coffee KC1&Coffee&Softs\\
ICE Cotton No 2 CT1&Cotton&Softs\\
ICE Orange Juice OJ1&Orange Juice&Softs\\
ICE Sugar No 11 SB1&Sugar&Softs\\
NYMEX Crude Oil CL1&Crude Oil (WTI)&Energy\\
NYMEX Heating Oil HO1&Heating Oil&Energy\\
NYMEX Natural Gas NG1&Natural Gas&Energy\\
NYMEX Palladium PA1&Palladium&Metals\\
NYMEX Platinum PL1&Platinum&Metals\\
\bottomrule \bottomrule
\end{tabular}
\end{table}

\begin{table}[htbp]
\centering
\caption{Results}
\label{tab2}
\footnotesize
\begin{tabular}{c|c|cc|cc|c}
\toprule \toprule
Commodity&$AE$&$D_{HW}$&$D_G$&$H_{LW}$&$H_{GPH}$&$EI$\\
\midrule
Cocoa&0.9728&1.4665&1.4605&0.3542&0.3367&0.1594\\
Coffee&0.9680&1.4948&1.4606&0.4575&0.4665&0.0469\\
Copper&0.8264&1.5613&1.4974&0.6205&0.6992&0.1843\\
Corn&0.9015&1.4592&1.4299&0.5241&0.4858&0.0744\\
Cotton&0.9564&1.4702&1.4564&0.5057&0.4735&0.0439\\
Crude Oil (Brent)&0.8919&1.5307&1.5084&0.5620&0.5986&0.0988\\
Crude Oil (WTI)&0.9427&1.5243&1.4987&0.5466&0.4499&0.0309\\
Feeder Cattle&0.3857&1.3498&1.3166&0.5751&0.3882&0.3500\\
Gold&0.5759&1.5161&1.4707&0.4278&0.4067&0.2277\\
Heating Oil&0.9568&1.4943&1.4916&0.5081&0.4592&0.0280\\
Lean Hogs&0.7081&1.3894&1.3584&0.3795&0.4256&0.2161\\
Live Cattle&0.4527&1.4206&1.3773&0.4433&0.4306&0.2985\\
Lumber&1.0040&1.4301&1.4428&0.4278&0.3603&0.1236\\
Natural Gas&1.1140&1.5246&1.4781&0.5210&0.5204&0.0607\\
Oats&0.9365&1.3926&1.3696&0.4105&0.2364&0.2152\\
Orange Juice&0.8770&1.4266&1.3899&0.4126&0.3399&0.1659\\
Palladium&1.0230&1.4266&1.4210&0.5625&0.5970&0.1109\\
Platinum&0.7443&1.4686&1.4845&0.5535&0.5465&0.1393\\
Rough Rice&0.8525&1.4278&1.4181&0.4512&0.4635&0.1149\\
Silver&0.8515&1.5161&1.4914&0.4685&0.4448&0.0861\\
Soybean Meal&0.8861&1.4448&1.4328&0.4878&0.4548&0.0884\\
Soybean Oil&0.7286&1.4735&1.4364&0.5330&0.5307&0.1465\\
Soybeans&0.7649&1.4900&1.4745&0.5266&0.5173&0.1209\\
Sugar&0.9759&1.4786&1.4818&0.5543&0.5505&0.0573\\
Wheat&0.9133&1.5129&1.4829&0.4626&0.5117&0.0453\\
\bottomrule \bottomrule
\end{tabular}
\end{table}

\end{document}